\documentclass[runningheads]{llncs}
\usepackage{graphicx}
\usepackage{subfig}
\usepackage[export]{adjustbox}
\usepackage{amsmath}
\usepackage{hyperref}
\usepackage[capitalize]{cleveref}
\usepackage{float}
\usepackage{multirow}
\graphicspath{{images/}}
\usepackage{tabularray}
\UseTblrLibrary{diagbox}
\usepackage{mathtools,xparse}

\newlength{\subcolumnwidth}

\newcommand{\nextsubcolumn}[1][]{%
  \cr\noalign{\hfill}
  \if\relax\detokenize{#1}\relax\else\hsize=#1\setlength{\subcolumnwidth}{\hsize}\fi
}

\begin{document}
\title{Defocus Blur Synthesis and Deblurring via Interpolation and Extrapolation in Latent Space}


\author{Ioana Mazilu*  \and
Shunxin Wang*     \and
Sven Dummer\and
Raymond Veldhuis  \and \\
Christoph Brune   \and
Nicola Strisciuglio  }
\authorrunning{I. Mazilu et al.}
%
\institute{University of Twente, Netherlands. \\
}

\maketitle              
\def\thefootnote{*}\footnotetext{These authors contributed equally to this work. Contact: s.wang-2@utwente.nl} 
\begin{abstract}
Though modern microscopes have an autofocusing system to ensure optimal focus, out-of-focus images can still occur when cells within the medium are not all in the same focal plane, affecting the image quality for medical diagnosis and analysis of diseases. We propose a method that can deblur images as well as synthesize defocus blur. We train autoencoders with implicit and explicit regularization techniques to enforce linearity relations among the representations of different blur levels in the latent space. This allows for the exploration of different blur levels of an object by linearly interpolating/extrapolating the latent representations of images taken at different focal planes. Compared to existing works, we use a simple architecture to synthesize images with flexible blur levels, leveraging the linear latent space. Our regularized autoencoders can effectively mimic blur and deblur, increasing data variety as a data augmentation technique and improving the quality of microscopic images, which would be beneficial for further processing and analysis. The code is available at \url{https://github.com/nis-research/linear-latent-blur}.

\keywords{Microscope images \and Deblurring \and Defocus blur synthesis \and Regularized autoencoders.}
\end{abstract}

\section{Introduction}  \label{sec:sec1}
Computer vision models have become increasingly popular in biomedical image processing, particularly with the advancement of deep learning techniques, leading to improved performance for tasks like cell segmentation and disease classification~\cite{Lugagne2020,Pandey2022,Santos2021}. However, image quality greatly impacts the performance of computer vision models. In the biomedical field, low-quality microscopy images can compromise image analysis and diagnosis. 

For instance, high-resolution cell images can be obtained using a confocal microscope. An autofocus component helps find the optimal focal plane for capturing a cell slide~\cite{Kumar2020}.
However, this task is often complicated by out-of-focus light, as not all cells are on the same focal plane and have thick structures. Thus, some cell images show less sharp regions due to out-of-focus areas, complicating the automated biomedical analysis~\cite{koab237}. 

Several deep-learning deblurring solutions have emerged in recent years to tackle this problem. They can be categorized into two groups: blur kernel estimation followed by deblurring~\cite{Quan2021} and kernel-free approaches~\cite{Nimisha2017,Wang2022,Liang2021,Zhang2022}. Quan et al.~\cite{Quan2021} proposed a non-blind deblurring network based on a scale-recurrent attention module. In \cite{Wang2022} and \cite{Basty2018}, the authors used multiscale U-net architectures for deblurring and image super-resolution tasks. These methods rely on local or global residual connections, which are useful for recovering information that may be lost through downsampling, as well as for optimizing the training process~\cite{Mao2016}. The authors of \cite{Lee2019} proposed a defocus map estimation model. The defocus map can be used to compute the pixel-wise blur level for blur enhancement and blur kernel estimation for deblurring. Jiang et al. \cite{Jiang2020} tackled multi-cause blur and proposed methods to recover sharp images from either motion or defocus blur. Zhang et al.~\cite{Zhang2022} reported state-of-the-art results for deblurring microscopic images using a CycleGAN-based model, which learns a reversible mapping between sharp and out-of-focus images. However, these methods entail high computational costs from the nature of the complex architectures and lack the flexibility of removing blur from images with defocus levels different from those seen during training.

\begin{figure*}[!t]
    \centering
\subfloat[$z_{0}$]{\includegraphics[height=1.9cm]{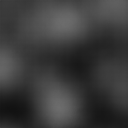}}
\hspace{2mm}
\subfloat[$z_{4}$]{\includegraphics[height=1.9cm]{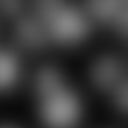}}
\hspace{2mm}
\subfloat[$z_{8}$]{\includegraphics[height=1.9cm]{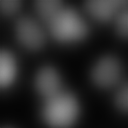}}
\hspace{2mm}
\subfloat[$z_{12}$]{\includegraphics[height=1.9cm]{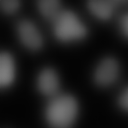}}
\hspace{2mm}
\subfloat[$z_{16}$]{\includegraphics[height=1.9cm]{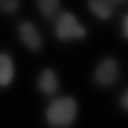}}
\vspace{-2mm}
    \caption{A nuclei-labeled cell slide captured at five focal lengths. A z-stack level (ranging from $z_0$ to $z_{16}$) indicates the level of blur. We enforce linearity in the latent space among image representations of different blur levels of one slide.}
   \label{fig:slides-intro}
\end{figure*}

In this paper, we propose a generative model that uses an autoencoder for both blur synthesis and deblurring. The unknown relation between latent representations of blur levels obtained with a vanilla autoencoder does not allow traversals of the latent space to generate images with lower or higher blur levels. We thus design training constraints that enforce a certain structure in the latent space, such as a linearity relation. We use a regular autoencoder as the baseline model and apply implicit and explicit regularization to enforce linearity among the image representations of a cell slide captured at different focal planes, such as those shown in \cref{fig:slides-intro}. The autoencoders are trained to synthesize defocus blur. Leveraging the enforced linearity, we can synthesize a blurry image by linearly interpolating the latent representations of two images of the same cell slide with different levels of blur. Further, the linear relation among blur levels enables synthesizing a sharper image by extrapolating representations of blurry images from the same cell slide. 
 
Our contributions are: 1) A model with a simple network architecture that serves as a versatile solution for both defocus blur synthesis and deblurring. 2) Adaptability to different blur levels enabling the recovery of in-focus images, even when the blur level of the reference images is unknown.

\begin{figure*}
\centering
\includegraphics[width=0.9\textwidth]{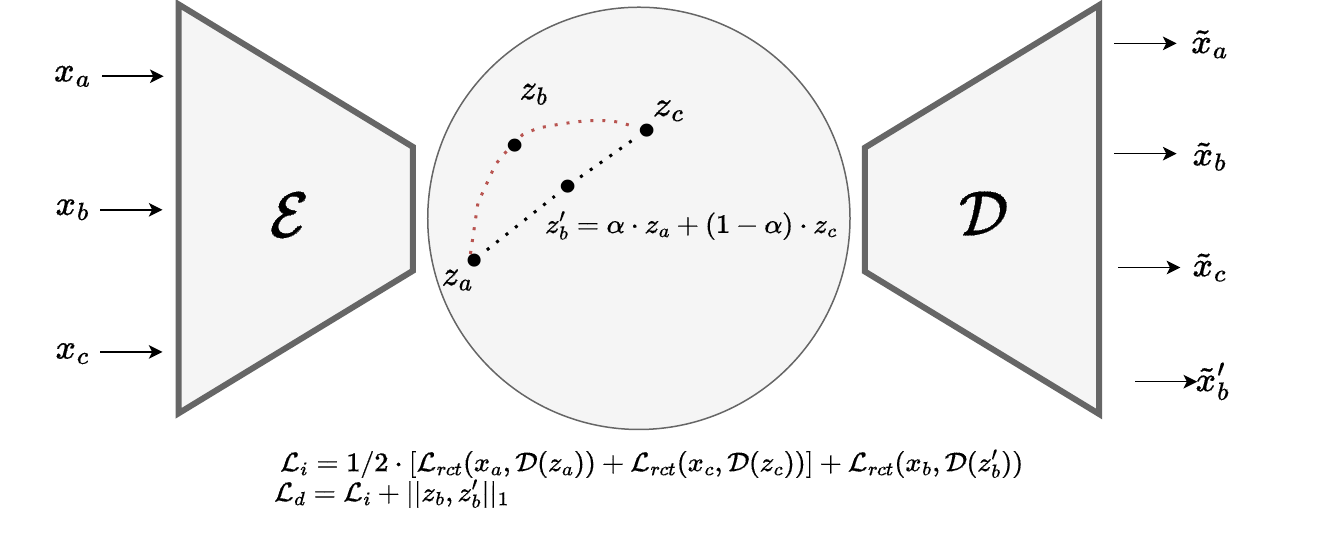}
\vspace{-2mm}
\caption{Given a triplet of inputs \{$x_a$, $x_b$, $x_c$\}, we generate their corresponding reconstructions $\tilde x_a$, $\tilde x_b$, $\tilde x_c$ and a synthetic blurry image $\tilde x'_b$ based on the linearly interpolated representation $z'_b$. $\mathcal{E}$ and $\mathcal{D}$ are an encoder and a decoder network.}
\label{fig:model_architecture}
\end{figure*}

\section{Proposed method} \label{sec:sec3}

\subsection{Imposing linearity onto latent space}
We hypothesize that a linear relation among latent representations of images with different blur levels taken from one cell slide allows us to generate images with flexible levels of blur. As shown in~\cref{fig:model_architecture}, given a triplet of images \{$x_a$, $x_b$, $x_c$\} captured at different focal lengths from the same cell slide, with an increasing blur level, we impose that their image representations follow the linear relationship in the latent space as:
\begin{equation}
    z'_{b} =  \alpha\cdot z_{a}+(1-\alpha)\cdot z_{c},
\end{equation}
\noindent where $z_{i} = \mathcal{E}(x_i)$ is the representation of the image $x_i$ computed with an encoder network $\mathcal{E}$, $z'_{b}$ is the latent representation interpolated from $z_a$ and $z_c$ and corresponds to image $x_{b}$,  and $\alpha$ is the interpolation parameter to control the level of blur. As $\alpha$ increases from 0 to 1, the level of blur decreases.

With the enforced linearity, we can synthesize the less blurry image $\tilde x'_a$, associated with $x_a$,  by extrapolating the latent representation $z'_a$ from the latent representations $z_b$ and $z_c$ of two images $x_b$ and $x_c$ ($x_b$ has a lower level of blur than that of $x_c$), as shown below:

\begin{equation}
    \tilde x'_{a} = \mathcal{D}(z'_a) = \mathcal{D}(\frac{1}{\alpha}\cdot z_{b} - \frac{1-\alpha}{\alpha}\cdot z_{c}) ,
\end{equation}

 \noindent where $\mathcal{D}$ is a decoder network and $z'_a$ is the extrapolated representation from $z_b$ and $z_c$. To achieve this, we train an autoencoder to reconstruct $x_a$ and $x_c$ from $z_a$ and $z_c$, and $x_b$ from $z'_b$. Extrapolation of latent representations is applied only in the test phase, and the linearity in the latent space affects directly the performance of deblurring. We apply indirect and direct regularization in the latent space to investigate how it affects image quality. 

\paragraph{Indirect regularization}
induces that the linearity is not directly embedded in the latent space, but is achieved by reconstructing the intermediate image $x_b$ using the interpolated representation of $z_a$ and $z_c$ without using $x_b$ as input. The objective function is:

\begin{equation} \label{loss_weak}
    \mathcal{L}_{i} = \frac{1}{2}\cdot[\mathcal{L}_{rct}(x_a,\mathcal{D}(z_a))+\mathcal{L}_{rct}(x_c,\mathcal{D}(z_c))] + \mathcal{L}_{rct}(x_b,\mathcal{D}({z'_b})),
\end{equation}
where the first term is the sum of the $L_1$ reconstruction losses of $x_a$ and $x_b$ using their corresponding learned latent representations and the second term is the $L_1$ reconstruction loss of the $x_b$ decoding from the interpolated representation $z_b'$.

\paragraph{Direct regularization}
adds a constraint that minimizes directly the $L_1$ distance between the interpolated latent representation $z'_b$ and the associated representation $z_b$ of image $x_b$, thus the objective function is: 
\begin{equation} \label{loss_strong}
    \mathcal{L}_{d} = \mathcal{L}_{i} + ||z_b-z'_b||_{1}.
\end{equation}
The indirect regularization may result in a latent space where interpolated latent representations are decoded into images visually similar to the real data, without forcing the non-interpolated latent codes to be linearly dependent~\cite{Sainburg2018}. The direct regularization explicitly ensures linearity in the latent space.

\subsection{Evaluation metrics} \label{ssec:evaluation}

The goal of this study is to model a latent space with a linear constraint, such that we can exploit interpolation and extrapolation to reconstruct images with flexible levels of blur. We evaluate the geometric properties of the latent space and image quality for blur synthesis and deblurring.

\paragraph{Linearity in latent space.}  \label{sssec:ld-metrics}

We quantify the degree of linear dependence among image representations based on two geometric properties. First, given three consecutive latent representations in terms of blur level, we measure their linearity based on the cosine similarity between the distance vectors obtained from each pair of neighbouring representations $z_n$ and $z_{n+1}$.

We call this the Linear Dependence Score ($\mathrm{LDS}$):
 \begin{equation} \label{apcs}
\mathrm{LDS} = \frac{1}{N-2}\sum_{n = 1}^{N-2}\frac{(z_{n-1}-z_n) \cdot (z_n-z_{n+1})}{||z_{n-1}-z_n||_2 \cdot ||z_n-z_{n+1}||_2},
\end{equation}

\noindent where N is the number of blur levels in the dataset, and $z_{n}$ is the latent representation of an image at blur level $n$. $\mathrm{LDS}$ ranges from -1 to 1, with higher values indicating a higher degree of compliance with the expected geometric property.

Second, since we traverse the latent space of a cell slide in fixed steps of $\frac{N}{\alpha}$, we assess whether the distance between neighbouring latent representations is equal between a pair of interpolated and a pair of non-interpolated image representations. To measure this property, we propose a metric called Average Pairwise Distance ($\mathrm{APD}$):
\begin{equation} \label{apd}
\mathrm{APD} = \frac{1}{N-1}\sum_{n=0}^{N-2} \frac{|d(z_{n}, z_{n+1}) - d(z'_{n}, z'_{n+1})|}{|d(z_{0}, z_{N-1})|},
\end{equation}
\noindent where $z_{n}$ and $z_{n+1}$ are latent representations of two consecutive images in terms of blur level. The score is normalized by the distance between the representations of the lowest and highest blur levels. $\mathrm{APD}$ ranges from 0 to 1 and a lower value indicates that the latent space approaches the desired structure. Moreover, visual inspection of the latent space is done by mapping the latent representations to a 2D space via PCA.

\paragraph{Image quality.}
We evaluate image quality  using a commonly used metric, Peak Signal-to-Noise-Ratio ($\mathrm{PSNR}$), 

\begin{equation}
    \mathrm{PSNR}_I^R = 20\cdot \log_{10}\frac{\max(I)}{\sqrt{\frac{1}{mn}\sum_{i=0}^{m-1}\sum_{j=0}^{n-1} (I(i,j) - R(i,j ))^2}}.
\end{equation}
This measures the similarity between the images $I$ and $R$. For instance, $\mathrm{PSNR}_{grd}^{extr_d}$ compares the deblurred image using an extrapolated latent representation to the corresponding ground truth sharp image. $\mathrm{PSNR}_{grd}^{b}$ compares the reconstructed blurry image with the ground truth blurry image. 

\section{Experiments and results} \label{sec:sec4}

\subsection{Dataset}

We use the BBBC006v1 collection obtained from the Broad Bioimage Benchmark Collection \cite{Ljosa2012}, which contains 384 cell slides stained with two markers to label the nuclei and structure of cells respectively. The sets of nuclei and cell structure images are noted as w1 and w2 sets. Each cell slide is captured at 34 focal lengths. In total, there are $384\times 2\times 34$ images. 
We only use the images captured above the optimal focal plane (z-stack$=$16) with even z-stack levels for both training and testing (z-stack$\le $16). We split it into training, validation, and testing sets, in a 7:1:2 ratio. All z-stack levels corresponding to one cell slide are assigned to the same set.
We use triplets of one slide captured at different focal lengths as input for the models. For the training phase, we use the triplets: ($z_a$, $z_b$, $z_c$) where $2b=a+c$ and $a$, $b$ and $c$ are even z-stack levels in the dataset. We only use even z-stack levels since changes between two consecutive blur levels (an even and an odd z-stack level) do not exhibit significant variation in the data.

\subsection{Architecture and training}
As the baseline model, we design an autoencoder with a simple architecture but can achieve image reconstruction marginally well. It has five convolutional layers in the encoder and six transposed convolutional layers in the decoder. Each convolutional layer consists of a two-strided convolution with a kernel size of $3 \times 3$, followed by batch normalization and Leaky ReLU activation. In the decoder, the structure is symmetrical, with transposed convolution replacing convolution operations. The last layer is a convolutional layer with kernel size $3 \times 3$, followed by a Sigmoid activation. The encoder layers have 64, 128, 256, 512, and 1024 filters. The models are trained for 40 epochs, with batch size 40. We use Adam optimizer with learning rate $10^{-4}$. We generate 10 crops of size 128×128 from each image (4 corner crops, 1 center crop, and their corresponding horizontally-flipped versions). Using the same architecture, the regularized models are trained with the proposed regularizations. We train models separately on the w1 and w2 sets, due to the difference in their data distributions.

\subsection{Results}
\paragraph{Linearity in the latent space.}

We show the 2D projections of latent representations of a set of images (from the same cell slide but captured at different focal lengths) in~\cref{fig:2d-projections}. The direct regularization forces the representations to be more clustered and arranged along a line. With indirect regularization, the distribution in the latent space of the representation of images with an increasing blur level is almost the same as that of the baseline model. We report the results on the linearity of the learned latent representations in~\cref{tab:results_linearity}, for the models trained on w1 and w2 sets, respectively. Direct regularization leads to substantial changes in the structure of the latent space. With direct regularization, interpolated or extrapolated latent representations lie closer in the latent space to their associated representations generated by the encoder. This means that images decoded from interpolated representations can be more similar to the images reconstructed from the latent representations of the real images, compared to those obtained with the other two models.

\hspace{-2.5em}
\begin{minipage}[t]{0.6\linewidth}
   \vspace{-2ex}
   \centering
   \includegraphics[width=0.8\linewidth]{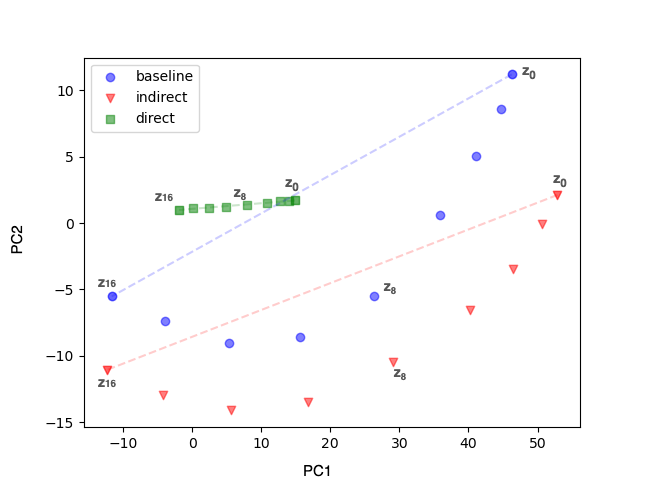}
   \captionof{figure}{2D latent representations of a cell slide.}
   \label{fig:2d-projections}
\end{minipage} 
\begin{minipage}[t]{0.4\linewidth}
\vspace{3ex}
\captionof{table}{Quantitative results of the linear dependence. The arrows indicate whether a lower or a higher score is desirable.}
\label{tab:results_linearity}
\SetTblrInner{rowsep=0pt}
\begin{tblr}[]{
      width = \linewidth,
      cells = {c},
      rows = {-1em,ht=2mm,font=\tiny},
      column{3-5} = {0.12\linewidth,c},
      column{2} = {0.12\linewidth},
     cell{2}{1} = {r=2}{0.07\linewidth} ,
     cell{4}{1} = {r=2}{0.07\linewidth} ,
      hline{2} = {2-5}{0.08em},
      hline{ 1,4,6} = {-}{0.08em},
      hline{3,5} = {1-5}{0.05em},
      vline{2} = {-}{},
      }
       \label{tab:results-ldsw1}
    \bfseries Data & \bfseries  Model  & \bfseries{Baseline} &\bfseries {Indirect} & \bfseries{Direct} \\
   \bfseries{  w1 set} &{$\mathrm{LDS}\uparrow$} & {0.59} & {0.62} & {\textbf{0.76}} \\
    &{$\mathrm{APD}\downarrow$ }&{ 0.036} & {0.028} & {\textbf{0.014}} \\
      \bfseries{w2 set} &{$\mathrm{LDS}\uparrow$}  & {0.52 }& {0.47} & {\textbf{0.75}}\\
        &{$\mathrm{APD}\downarrow$} & {0.037} & {0.048} & {\textbf{0.031}}\\
        \end{tblr}\\ 
\end{minipage}

\begin{figure}
\centering
 \subfloat[]{\includegraphics[height=4.cm]{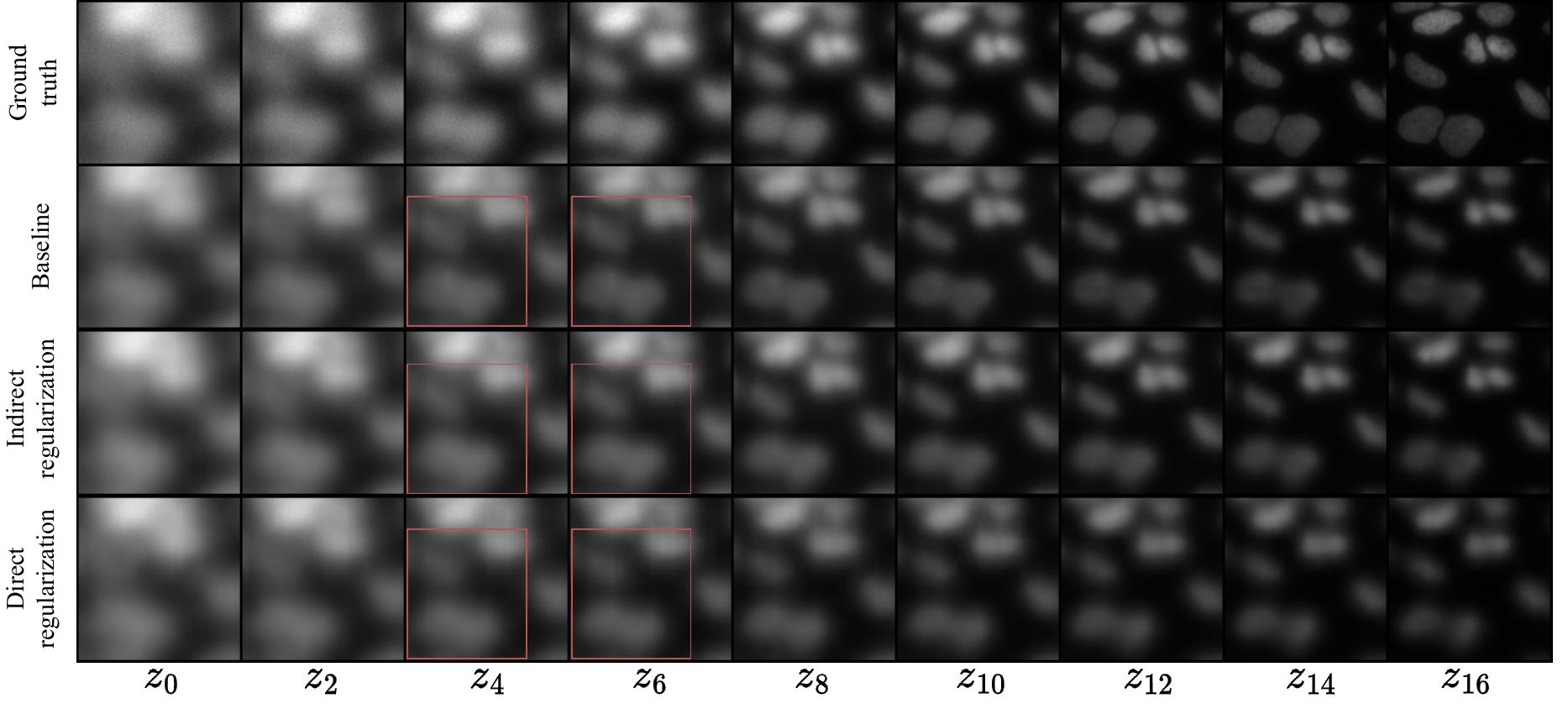}\label{fig:w1_blur}} \hspace{2em}
 \subfloat[]{\includegraphics[height=4.cm]{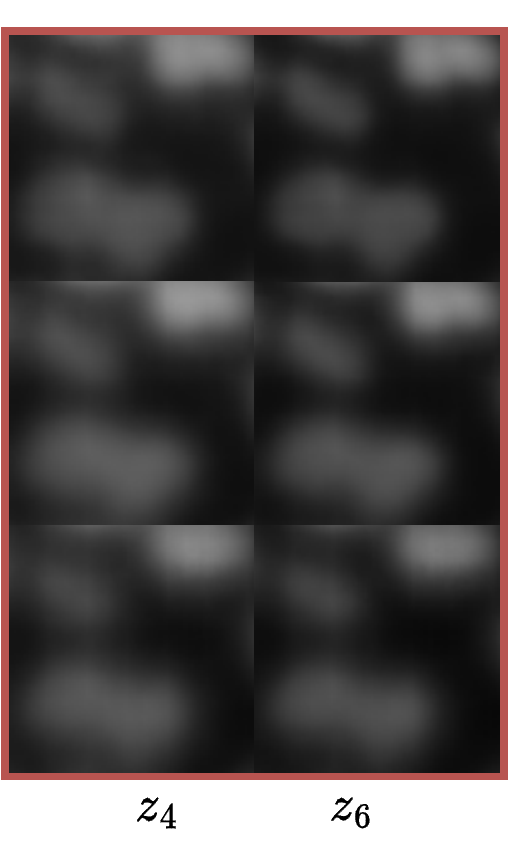}\label{fig:w1_blending}} \\
\subfloat[]{\includegraphics[height=3.8cm]{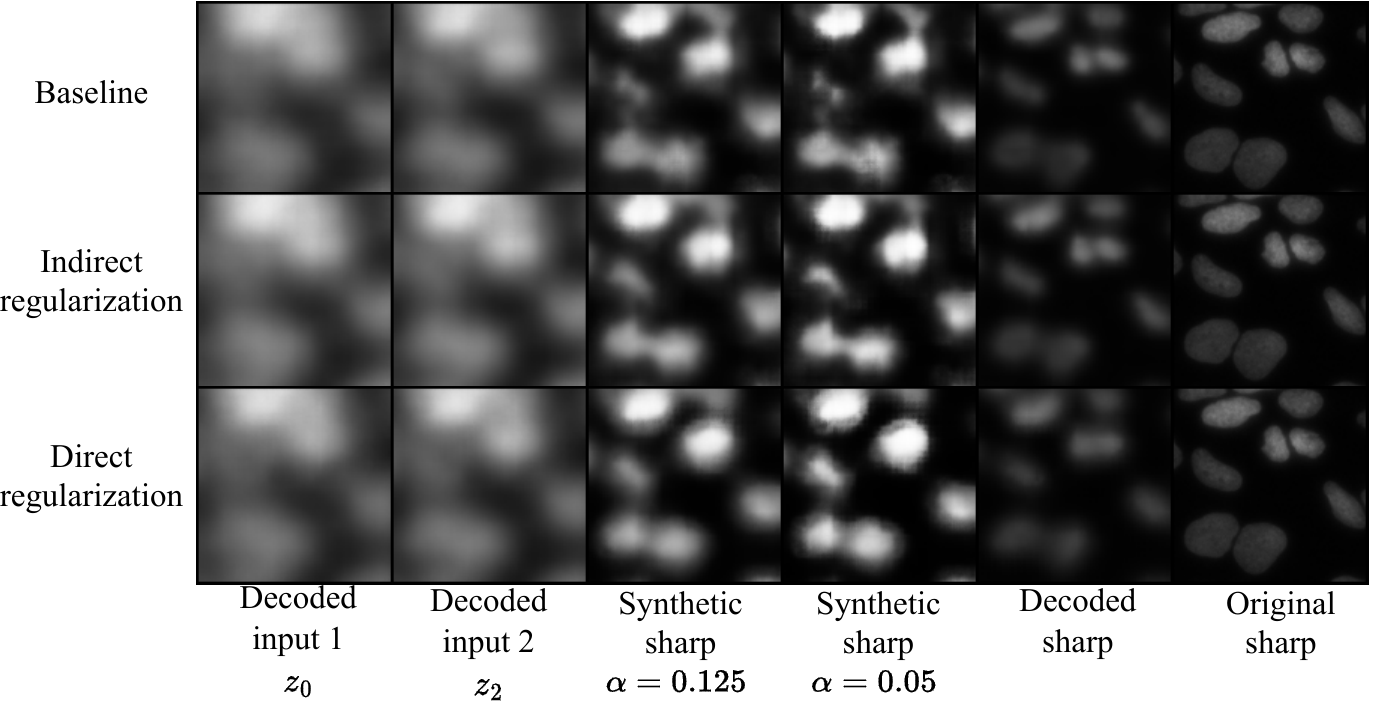}\label{fig:w1_deblur_by_models}}
\caption{(a) Synthesized blur for a nuclei-labeled slide using the baseline, indirectly and directly regularized models. Each row contains images with the blur level transitioning from z-stack 0 (left) to z-stack 16 (right), (b) Zoom-in view of the area within the frame in (a) highlights the blending effect by the baseline, (c) Example of image deblurring using 3 models. Synthetic sharp images are obtained through linear extrapolation between representations of two slides with z-stack levels 0 and 2, using different values for $\alpha$. }
\end{figure}
 
 \begin{table}
\centering
\caption{Results of the baseline model, and the directly- and indirectly-regularized models, on the w1 and w2 sets. Blur synthesis and deblurring are evaluated. The arrows indicate whether a lower or a higher score is better. The best scores are highlighted.}
\label{tab:results-quality}
\resizebox{\linewidth}{!}{%
\begin{tblr}{
  cells = {c},
  cell{2}{3} = {c=3}{0.283\linewidth},
  cell{2}{6} = {c=3}{0.283\linewidth},
  cell{1}{2} = {r=2}{},
  cell{3}{1} = {r=2}{},
  cell{5}{1} = {r=2}{},
  column{1} = {l},
  column{2} = {l},
  row{1,2} = {font=\bfseries},
  hline{1,7} = {-}{0.08em},
  hline{2} = {3-8}{0.05em},
  hline{3,5} = {-}{0.05em},
  hline{4,6} = {2}{l},
  hline{4,6} = {3-7}{},
  hline{4,6} = {8}{r},
  vline{3} = {1-6}{},
  vline{2} = {1-6}{},
  vline{6} = {1-6}{}
}
 & \diagbox{ Metric}{Model} & Baseline & Indirect & Direct & Baseline & Indirect & Direct\\
 Experiment &  & w1 set &  &  & w2 set &  & \\
Blur synthesis & $\mathrm{PSNR}_{b}^{interp_b}$ $\uparrow$ & \textbf{31.97} & 31.77 & 31.60 & 32.35 & 32.48 & \textbf{33.05}\\
 & $\mathrm{\mathrm{PSNR}}_{grd}^{interp_b}$ $\uparrow$ & 29.09 & \textbf{31.03} & 28.30 & \textbf{29.78} & 29.46 & 29.02\\
Deblurring & $\mathrm{\mathrm{PSNR}}_{d}^{extr_d}$ $\uparrow$ & \textbf{23.89} & 23.02 & 23.52 & \textbf{22.47} & 22.22 & 22.03\\
 & $\mathrm{PSNR}_{grd}^{extr_d}$ $\uparrow$ & 22.55 & \textbf{22.60} & 21.98 & 24.00 & 23.49 & \textbf{24.46}
\end{tblr}
}
\end{table}

\paragraph{Blur synthesis.}

 In~\cref{tab:results-quality}, we report the comparison of the quality of images reconstructed using interpolated representations, against reconstructed images using the representations associated with ground truth blurry images ($\mathrm{PSNR}_{b}^{interp_b}$) and ground truth blurry images ($\mathrm{PSNR}_{grd}^{interp_b}$).
 We show the synthesized blurry images in \cref{fig:w1_blur}. With the baseline model, reconstructions from linear traversals of the latent space between two points result in visually similar images compared with the ground truth blurry images. However, the reconstructed images show a blending effect between the two source images, rather than a reliable estimation of defocus blur effect, as shown in \cref{fig:w1_blending}. The visual quality of the synthetic blur improves with the addition of regularization, which helps to reduce the blending effect.

\paragraph{Deblurring.}

 We report the results of the quality of the deblurred images in~\cref{tab:results-quality}. We show examples of deblurred images of a slide in w1 set by the baseline and regularized models in \cref{fig:w1_deblur_by_models}. Using two blurry images, we can generate a sharper image. For the w1 set, the indirectly regularized autoencoder outperforms the baseline model when we compare the deblurred images with the reconstructed sharp images (see $\mathrm{PSNR}_{d}^{extr_d}$). For the w2 set, direct regularization performs the best. We observe that there is a trade-off between the desired geometric property and the image quality when applying direct regularization. With a better regularized latent space, the reconstructed image fidelity decreases slightly, while allowing to reconstruct and generate new images with different levels of blur using linear interpolation and extrapolation of the latent representations, respectively. To account for the clustering induced by the direct regularization, we also generate synthetic sharp images with an adjusted value for $\alpha$ ($\alpha=0.05$). We notice that this set of images shows slightly more sharpness compared to those using $\alpha=0.125$. This indicates that the levels of blur are indeed encoded along the linear direction in the latent space.

\begin{figure}
\centering
\includegraphics[height=4.8cm]{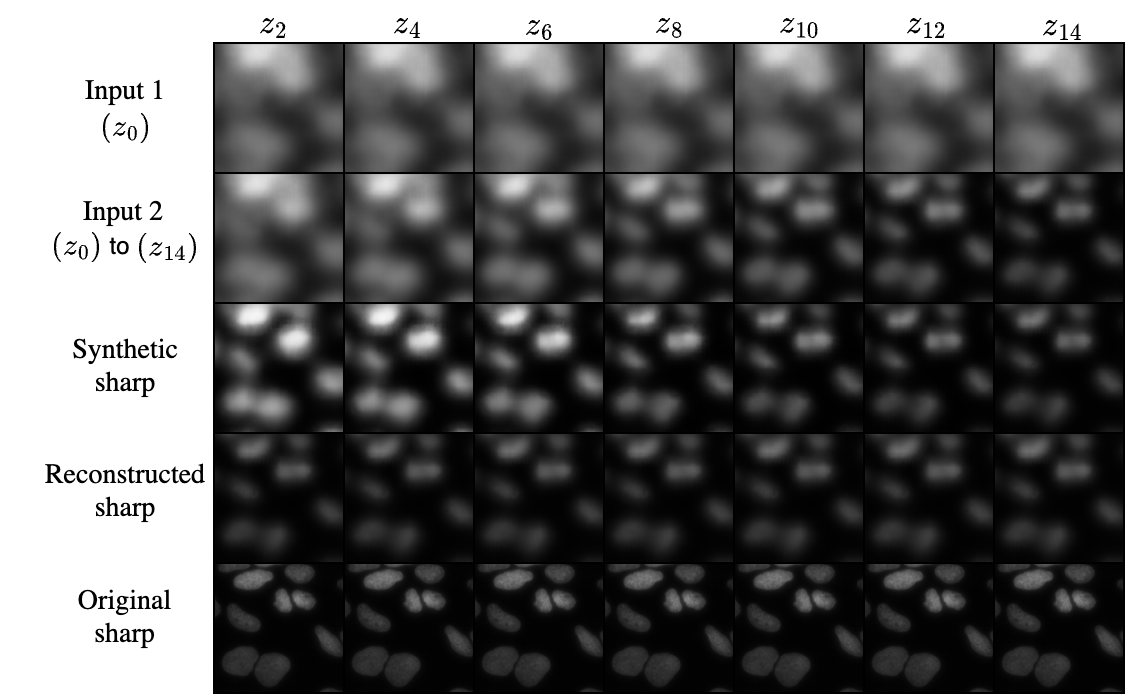}
\caption{Effect of the blur level in the input images on the synthetic sharp image, when the optimal interpolation parameter $\alpha$ is known, using the indirectly-regularized model.}
\label{fig:deblur-progression}
\end{figure}

With our regularized model, even when the sharp image is generated from two images with high levels of blur, a considerable level of detail is recovered. \cref{fig:deblur-progression} shows how the blur level of input images affects the deblurring process. We fix one image at z-stack 0 and vary the other one from z-stack 2 to z-stack 14. These results are in line with those from a similar study \cite{Luo2021}, where the level of detail recovered in the deblurred images decreases with an increase in the focal plane at which slides are captured.

\section{Discussion and future work} \label{sec:sec5}

Our results suggest the feasibility of blur synthesis and deblurring through linear interpolation and extrapolation in the latent space. Imposing linearity onto the latent space enables us to control the level of blur in an image by interpolating or extrapolating representations.
With a simple architecture, we achieve a versatile solution for blur synthesis and deblurring, while other works are usually limited to one application. 

The linear latent space enables the recovery of in-focus images, even when the blur level of the two reference images is unknown. One can dynamically adjust the value of $\alpha$ until reaching the optimal point. Besides, given a single blurry image as input, we can generate a second blurry image on top of it with a blur kernel, to obtain a deblurred in-focus image. 

From the curvilinear trajectory demonstrated in the 2D projections of the latent representations, we conjecture that there may be two directions in the latent space, one corresponding to blur levels and the other corresponding to image content. We suggest future work on disentanglement representation learning, i.e.  the representations of blur levels and image content are disentangled. This may allow for more precise reconstructions of deblurred images.

\section{Conclusions} \label{sec:sec6}

In this paper, we investigated the feasibility of models for both defocus blur synthesis and deblurring, based on linear interpolation and extrapolation in latent space. 
We enforce linearity among the representations of images of the same cell slide with different levels of blur, by indirect and direct regularization in the latent space. Therefore, linearly interpolating or extrapolating the representations of two differently blurred images (from the same cell slide) results in a meaningful representation that maps to an image with another level of blur. Our results show that the regularized models perform well on both blur synthesis and deblurring. The direct regularization results in a more linear latent space compared to a regular autoencoder, enabling a more precise mapping between extrapolated representations and their non-extrapolated versions.

\noindent\textbf{Acknowledgement}
This work was supported by the SEARCH project, UT Theme Call 2020, Faculty of Electrical Engineering, Mathematics and Computer Science, University of Twente.

\bibliographystyle{splncs04}
\bibliography{mybibliography}

\end{document}